\documentclass[12pt]{iopart}
\usepackage{iopams}
\usepackage{pstricks}
\usepackage{epsfig}
\begin{document}
\title[Quasi-long-range ordering in a finite-size $2D$ Heisenberg model]{
Quasi-long-range ordering in a finite-size $2D$ Heisenberg model
}

\author{O.\ Kapikranian$^{1,2}$, B.\ Berche$^{2}$, and
 Yu.\ Holovatch$^{1,3}$}

\address{$^{1}$
Institute for Condensed Matter Physics, National Acad. Sci. of Ukraine, UA-79011 Lviv, Ukraine}

\address{$^{2}$ Laboratoire de Physique des Mat\'eriaux, UMR CNRS 7556,
  Universit\'e Henri Poincar\'e, Nancy 1, F-54506  Vand\oe uvre les Nancy Cedex, France}

\address{$^{3}$
Institute f\"ur Theoretische Physik, Johannes Kepler Universit\"at
Linz, A-4040 Linz, Austria}

\begin{abstract}
We analyse the low-temperature behaviour of
the Heisenberg model on a two-dimensional lattice {\em of
finite size}. Presence of a residual magnetisation in a finite-size system enables
us to use the spin wave approximation, which is known to give reliable results for
the $XY$ model at low temperatures $T$. For the system considered, we find that 
the spin-spin correlation function decays as $1/r^{\eta(T)}$ for large separations 
$r$ bringing about  presence of a quasi-long-range ordering. We give analytic estimates
for the exponent $\eta(T)$ in different regimes and support our findings by Monte Carlo 
simulations of the model on lattices of different sizes at different temperatures.
\end{abstract}

\eads{\mailto{akap@icmp.lviv.ua},
\mailto{berche@lpm.u-nancy.fr},
\mailto{hol@icmp.lviv.ua}}

\maketitle

The long history of the Heisenberg model in two dimensions is
characterised by a competition between two contrary opinions about
properties of this model. The early observation, made by
Peierls~\cite{Peierls35}, about long-wavelength lattice waves
destroying the localisation of particles on their lattice sites in
two-dimensional crystals was followed later by a similar result
for $2D$ magnets of continuous symmetry where the spontaneous
magnetisation is destroyed by long-wavelength spin waves. Proven
mathematically by Mermin and Wagner~\cite{MerminWagner66}, this
fact denied the very possibility of a ferromagnetic phase
transition in this type of systems. Although the high-temperature
series for the Heisenberg and $XY$ models in $2D$, presented by
Stanley and Kaplan~\cite{StanleyKaplan} approximately at the same
time, gave indication of a phase transition in both models.
Being qualitatively similar in those works, the two models had
quite different developments afterwards. The $2D$ $XY$ model has
become famous for the Berezinskii-Kosterlitz-Thouless (BKT)
transition~\cite{KosterlitzThoulessBerezinskii} to a
quasi-long-range ordered (QLRO) phase. This special type of
ordering cannot be characterised by an order parameter in an
infinite system and manifests in a power law decay of the
spin-spin correlation function with distance. The low-temperature
properties and critical behaviour of the $XY$ model on a $2D$
lattice are governed by interactions between topological defects
which appear in the system~\cite{Nelson02,Chaikin95,Kenna06}. This
scenario is deeply connected to the symmetry of the model. In the
$XY$ model rotations of a spin form an Abelian group what allows
formation of stable topological defects like spin vortices and others, in
this sense it can be called an Abelian model in contrast to the
non-Abelian ones. The Heisenberg model is non-Abelian, this is the
main reason to deny a possibility of a BKT transition in it, since
no stable topological defects (instantons) can be formed. The crucial evidence
for an absence of a phase transition in the $2D$ Heisenberg model
came from the renormalization treatment made by
Polyakov~\cite{Polyakov75,Izyumov88}. Now it is commonly believed that this
model does not exhibit any phase transition at non-zero
temperatures, although there are still some controversies (see e.g.
Refs.~\cite{Niedermayer97,Patrascioiu95,Patrascioiu02,Patrascioiu01,PatrascioiuSeiler}) 
and reports were made 
that disagree with the outcome of Polyakov's work, assuming the possibility of a
phase transition at finite temperature in the $2D$ Heisenberg model,
similar to the BKT transition in the $2D$ $XY$ model (see \cite{Kenna06} and 
references therein). The last question is important in the context of an
asymptotic freedom of QCD at $4D$ \cite{Izyumov88,Kogut79}.

The discussion above concerns {\em infinite systems}. And it is an
infinite $2D$ model of continuous symmetry for which the
Mermin-Wagner theorem has been proven. However, either in Monte
Carlo simulations or even in reality we always deal with finite
physical systems. It is well known now that in a finite $2D$
$XY$-spin system below the BKT transition temperature there is a
non-vanishing magnetisation which tends to zero only in the
thermodynamic limit~\cite{TobochnikChester79,Bramwell&Co}.
This observation goes back to Berezinskii himself and is supported
by experimental measurements (see Ref.~\cite{BramwellHoldsworth}).

Although there is still no definite answer for the question
whether the $2D$ Heisenberg model can pass to a QLRO phase or not,
it is reasonable to assume that this model considered on a finite
lattice will certainly possess some ordering, i. e. non-vanishing
magnetisation, at low temperatures similar as the $2D$ $XY$ model
does. This assumption is clear from the obvious fact that in a
finite system, transition to the ordered phase at $T=0$ (when all
spins of the Heisenberg model are pointed in the same direction)
must be continuous. Hence we must see an appearence of some ordering as the
temperature approaches zero. This is confirmed by MC simulations
on the Heisenberg model in two dimensions~\cite{JMMM}.

Due to the above reasonings, we assume all spins ${\bf S}_{\bf
r}=(S^x_{\bf r}, S^y_{\bf r}, S^z_{\bf r})$ in the Hamiltonian of
the Heisenberg model on a two-dimensional lattice with spacing $a$
and sites defined by a vector ${\bf r}$:

\begin{equation}\label{Hamilton_1}
H = -\frac{1}{2}\sum_{\bf r}\sum_{\bf r'}J({\bf r-r'})\left(S^x_{\bf r}S^x_{\bf r'}
+S^y_{\bf r}S^y_{\bf r'}+S^z_{\bf r}S^z_{\bf r'}\right)\ ,
\end{equation}
being pointed approximately in the same direction at low enough
temperatures. This allows us to treat the model by means of the
spin-wave approximation (SWA). We consider the case of the nearest
neighbours interaction potential $J({\bf r-r'})=J\delta_{|{\bf
r-r'}|, a}$, and 1/2 stands in (\ref{Hamilton_1}) to prevent
double count of each bond.

\begin{figure} [th]
  \centerline{\includegraphics[width=9.5cm]{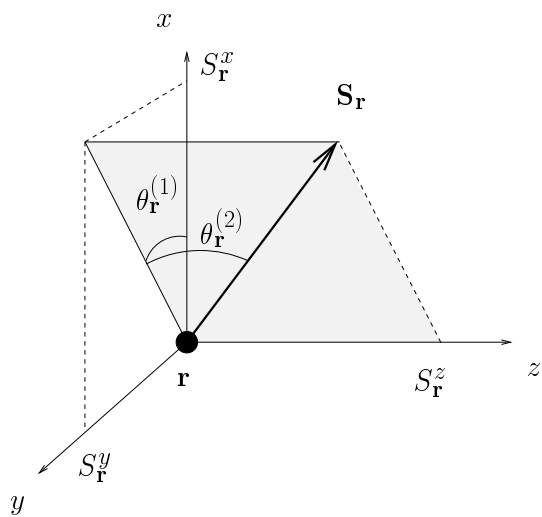}}
  \vspace*{-21ex}
        \caption{The angle variables $\theta^{(1)}_{\bf r}$ and $\theta^{(2)}_{\bf r}$
   that can be used to define the position of the spin ${\bf S}_{\bf r}$ placed at
   the site ${\bf r}$.}
   \label{Fig1}\end{figure}

We choose a special system of the angle coordinates
$\theta^{(1)}_{\bf r}$, $\theta^{(2)}_{\bf r}$ (see Fig.1),
defined by the relations:

\begin{eqnarray}\nonumber
S^x_{\bf r} &=& \cos\theta^{(1)}_{\bf r}\cos\theta^{(2)}_{\bf r}\ ,
\\\nonumber
S^y_{\bf r} &=& \sin\theta^{(1)}_{\bf r}\cos\theta^{(2)}_{\bf r}\ ,
\\\label{Angl_Var_Def}
S^z_{\bf r} &=& \sin\theta^{(2)}_{\bf r}\ ,
\end{eqnarray}
with $-\pi<\theta^{(1)}<\pi$,
$-\frac{\pi}{2}<\theta^{(2)}<\frac{\pi}{2}$. Note that the 
variables chosen are just slightly modified angles $\varphi$, $\theta$ of
the spherical coordinates: $\theta^{(1)}_{\bf r}=\varphi$,
$\theta^{(2)}_{\bf r}=\theta - \pi/2$. 
Assuming that the angles $\theta^{(1)}_{\bf
r}$, $\theta^{(2)}_{\bf r}$ are small at
low temperatures for all spins of the system, this choice of coordinates
enables us to make use of the SWA and to substitute the scalar product 
of two spins that stands in (\ref{Hamilton_1}):
\begin{eqnarray}\nonumber
S^x_{\bf r}S^y_{\bf r'}+S^y_{\bf r}S^y_{\bf r'}+S^z_{\bf r}S^z_{\bf r'} &=&
\cos(\theta^{(1)}_{\bf r}-\theta^{(1)}_{\bf r'})\cos(\theta^{(2)}_{\bf r}
-\theta^{(2)}_{\bf r'})
\\\label{Scal_Prod}
&+&\left(1-\cos(\theta^{(1)}_{\bf r}-\theta^{(1)}_{\bf r'})\right)\sin
\theta^{(2)}_{\bf r}\sin\theta^{(2)}_{\bf r'}\ ,
\end{eqnarray}
by an expression quadratic in $\theta^{(1)}_{\bf r}$, $\theta^{(2)}_{\bf r}$. 
Thus, (\ref{Scal_Prod}) can be written in the SWA as
\begin{equation}\label{SWA}
S^x_1S^x_2+S^y_1S^y_2+S^z_1S^z_2 \approx
1 - \frac{1}{2}(\theta_1^{(1)}-\theta_2^{(1)})^2
-\frac{1}{2}(\theta_1^{(2)}-\theta_2^{(2)})^2\ ,
\end{equation}
and the Hamiltonian (\ref{Hamilton_1}) is reduced to

\begin{equation}\label{Hamilton_2}
H = H_0 + H_1^{XY}(\{\theta^{(1)}\}) + H_1^{XY}(\{\theta^{(2)}\})\ ,
\end{equation}
where
\begin{equation}\label{Hamilton_XY}
H_1^{XY}(\{\theta\}) = \frac{1}{4}\sum_{\bf r}\sum_{\bf r'}J({\bf r-r'})
\left(\theta_{\bf r}-\theta_{\bf r'}\right)^2
\end{equation}
is the Hamiltonian of the $2D$ $XY$ model on the same lattice taken in the
SWA~\cite{Wegner67}. $H_0$ can be regarded as a shift in the energy scale.

The spin-spin correlation function with the assumption about
smallness of all $\theta^{(1)}_{\bf r}$, $\theta^{(2)}_{\bf r}$
reads:

\begin{equation}\label{Corr_Funct}
G_2(R) = \left<{\bf S_{\bf r}\cdot S_{\bf r+R}}\right> \approx
\left<\cos(\theta^{(1)}_{\bf r} - \theta^{(1)}_{\bf
r+R})\cos(\theta^{(2)}_{\bf r} - \theta^{(2)}_{\bf r+R})\right>\ ,
\end{equation}
where the angular brackets stand for the thermodynamic averaging:

\begin{displaymath}
\left<...\right> = \frac{1}{Z}{\rm Tr\ \!}(...\e^{-\beta
H}),\qquad\textrm{with}\quad Z = {\rm Tr\ \!} \e^{-\beta H}
\end{displaymath}
and
\begin{equation}\label{Spur}
{\rm Tr\ \!} ... = \prod_{\bf
r}\frac{1}{4\pi}\int_{-\pi}^{\pi}d\theta^{(1)}_{\bf r}
\int_{-\pi/2}^{\pi/2}d\theta^{(2)}_{\bf r}\cos\theta^{(2)}_{\bf
r}...\ .
\end{equation}

In the $2D$ $XY$ model the power law decay of the spin-spin
correlation function with distance serves as an indication of
QLRO. A suitable quantity to characterise the decay of the
correlation function with increase of the distance $R$ is the
temperature dependent exponent:
\begin{equation}\label{Eta_Def}
\eta(T) = -\lim_{R\to\infty}
\frac{\ln G_2(R)}{\ln R} .
\end{equation}
In the case of the $2D$ $XY$ model the SWA gives for the exponent
$\eta^{XY}$~\cite{Wegner67}:

\begin{equation}\label{EtaXY}
\eta^{XY}=1/(2\pi\beta J)\ ,
\end{equation}
that is reliable for small temperatures \cite{Berche}.

Due to the separation of the angle variables $\theta^{(1)}_{\bf
r}$, $\theta^{(2)}_{\bf r}$ in (\ref{Hamilton_2}) and
(\ref{Corr_Funct}) we can write for the correlation function:

\begin{equation}\label{G1xG2}
G_2(R)\ =\ G^{(1)}_2(R)\times G^{(2)}_2(R)\ ,
\end{equation}
where

\begin{equation}\label{G1}
G^{(1)}_2(R)= \frac{1}{Z_1}(4\pi)^{-N}\left(\prod_{\bf r'}\int_{-\pi}^{\pi}d
\theta^{(1)}_{\bf r'}\right)\e^{-\beta H_1^{XY}(\{\theta^{(1)}\})}
\cos(\theta^{(1)}_{\bf r} - \theta^{(1)}_{\bf r+R})
\end{equation}
and

\begin{equation}\label{G2}
G^{(2)}_2(R)=\frac{1}{Z_2}\left(\prod_{\bf r'}\int_{-\pi/2}^{\pi/2}d
\theta^{(2)}_{\bf r'}\cos\theta^{(2)}_{\bf r'}\right)\ \e^{-\beta H_1^{XY}
(\{\theta^{(2)}\})}\cos(\theta^{(2)}_{\bf r} - \theta^{(2)}_{\bf r+R})\ .
\end{equation}
$Z_1$ and $Z_2$ respectively originate from the integration over
$\theta^{(1)}_{\bf r}$ and $\theta^{(2)}_{\bf r}$ in the partition
function $Z=Z_1Z_2$. Although it  may be believed that the SWA applied to  $O(n)$
models automatically leads to $(n-1)\eta^{XY}$ exponent, the presence of the cosine 
in the integration element in (\ref{G2}) makes the problem more involved.

Now, to define the decay of the spin-spin correlation function
$G_2(R)$ for large distances $R$ it is enough to find the
asymptotic behaviour of $G^{(1)}_2(R)$ and $G^{(2)}_2(R)$ in the
limit $R/a\rightarrow\infty$. It is easy to see that
$G^{(1)}_2(R)$, Eq.(\ref{G1}), is just the correlation function of
the $2D$ $XY$ model, $G^{XY}_2(R)$, the asymptotic behaviour of which is
well known:

\begin{equation}\label{G_XY}
G_2^{(1)}(R) = G^{XY}_2(R)\ \approx\ (R/a)^{-\eta^{XY}}
\end{equation}
with $\eta^{XY}$ given by (\ref{EtaXY}). So, the problem is to
evaluate $G_2^{(2)}(R)$. We will follow the same scheme that has
been used to find $G^{XY}_2(R)$ \cite{Wegner67}. For this purpose
we pass to the Fourier variables:

\begin{equation}\label{Fourier}
\theta^{(2)}_{\bf r}=\frac{1}{\sqrt{N}}\sum_{\bf k}\e^{i{\bf
kr}}\theta_{\bf k}, \quad \theta_{\bf
k}=\frac{1}{\sqrt{N}}\sum_{\bf r}\e^{-i{\bf kr}}\theta^{(2)}_{\bf r},
\end{equation}
where $N$ is the total number of lattice sites and {\bf k} spans
the 1st Brillouin zone. Then the Hamiltonian (\ref{Hamilton_XY})
reads

\begin{displaymath}
H_1^{XY}(\{\theta\}) = J\sum_{{\bf k}\neq 0}\gamma_{\bf k}
\theta_{\bf k}\theta_{\bf -k}
\end{displaymath}
with $\gamma_{\bf k}\equiv 2-\cos k_xa-\cos k_ya$,
and the integration in (\ref{G2}) must be changed to

\begin{equation}\label{Spur_Four}
\left(\prod_{{\bf k}\in B/2}\int_{-\infty}^{\infty}d\theta^c_{\bf k}
\int_{-\infty}^{\infty}d\theta^s_{\bf k}\right)\
\end{equation}
with the product taken over a half of the 1st Brillouin zone
denoted by $B/2$. The cosine $\cos(\theta^{(2)}_{\bf r} -
\theta^{(2)}_{\bf r+R})$ in (\ref{G2}) can be presented in the
Fourier variables as

\begin{displaymath}
Re\ \exp\left[\frac{i}{\sqrt{N}}\sum_{{\bf k}\neq 0}(\eta^c_{\bf
k}\theta^c_{\bf k} +\eta^s_{\bf k}\theta^s_{\bf k})\right]
\end{displaymath}
with $\eta^c_{\bf k}\ =\ \cos{\bf kr}-\cos{\bf k(r+R)}$ and
$\eta^s_{\bf k}\ =\ -(\sin{\bf kr}-\sin{\bf k(r+R)})$. The product
of cosines in (\ref{G2}) can be expressed in the SWA as

\begin{equation}\label{Jacobian}
\prod_{\bf r}\cos\theta^{(2)}_{\bf r}\approx\prod_{\bf r}\ e^{-\frac{1}{2}
(\theta^{(2)}_{\bf r})^2}
= \exp\left[-\frac{1}{2}\sum_{\bf r}(\theta^{(2)}_{\bf r})^2\right]\ ,
\end{equation}
using the equality $\sum_{\bf r}(\theta^{(2)}_{\bf r})^2 =
\sum_{{\bf k}\neq 0}\theta_{\bf k}\theta_{\bf -k}$ we obtain an
integrable form for $G_2^{(2)}(R)$. After simple integration we
have

\begin{equation}\label{G2_Sum}
G^{(2)}_2(R) = \exp\left(-\frac{1}{\beta JN}\sum_{\bf k\neq
0}\frac{\sin^2\frac{\bf kR}{2}} {\gamma_{\bf k}+\frac{1}{2\beta
J}}\right)\ .
\end{equation}

Recall that we are interested in the long-distance behaviour of
the pair correlation function (\ref{Corr_Funct}) at low $T$. We
estimate the asymptotic behaviour of (\ref{G2_Sum}) in the limit
$R/a\rightarrow\infty$, $\beta J\rightarrow\infty$ for two cases:

\begin{equation}\label{Two_Cases}
G^{(2)}_2(R) \sim \left\{ \begin{array}{ll} (1+4\pi\beta
J)^{-\frac{1}{2\pi\beta J}}\quad &
\textrm{for}\quad\frac{(R/a)^2}{4\beta J}\gg 1\ ;\\
(R/a)^{-\frac{1}{2\pi\beta J}}\quad &
\textrm{for}\quad\frac{(R/a)^2}{4\beta J}\ll 1\ .
\end{array} \right.
\end{equation}
Thus, it is either constant with respect to $R$ or equivalent to
(\ref{G_XY}) depending on the asymptotic value of the ratio
$\frac{(R/a)^2}{4\beta J}$.

Substituting (\ref{Two_Cases}) into (\ref{G1xG2}) with the known
expression for $G_2^{(1)}(R)$, Eq.(\ref{G1}), we get the following
asymptotic behaviour of the spin-spin correlation function:

\begin{equation}\label{2EtaXY}
G_2(R) \sim \left(R/a\right)^{-2\eta^{XY}}\quad \textrm{for} \quad
\frac{(R/a)^2}{4\beta J}\ll 1\ ,
\end{equation}

\begin{equation}\label{1EtaXY}
G_2(R) \sim \left(R/a\right)^{-\eta^{XY}}\quad \textrm{for}\quad
\frac{(R/a)^2}{4\beta J}\gg 1\ .
\end{equation}

Let us clarify which of the two above estimates corresponds to the
behaviour observable in practice. An approach used in the above
derivations was based on the assumptions about smallness of the
temperature and finiteness of the lattice. Therefore, the limit
$\beta J\rightarrow\infty$ is physically grounded, because the lower
temperature we consider, the more likely ordering in the system
is. But the limit $R/a\rightarrow\infty$ as well as
$N\rightarrow\infty$ used to obtain approximate estimates of the
integrals in practice is limited by finiteness of the lattice:
$(R/a)^2 < N$. As the lattice size grows to infinity the ordering
disappears and our approach may become invalid. From the above
arguments we conclude that for a system of a {\em finite size} in the
low-temperature limit the estimate  (\ref{2EtaXY}) holds. Moreover, power-law 
asymptotics (\ref{2EtaXY}), (\ref{1EtaXY}) brings about QLRO present in the system.
Recall that these formulas were obtained by means of the SWA. Applicability of the
latter to $2D$ Heisenberg model has been justified at the beginning of this paper 
by a finite system size that leads to residual magnetisation at low $T$. Therefore, 
although the QLRO was not considered as an intrinsic property of a finite-size 
system, the power-law  asymptotics (\ref{2EtaXY}), (\ref{1EtaXY}) gives arguments 
in favour of its presence. 

Let us quote another argument that favours the QLRO in the $2D$ Heisenberg model, 
which by no means is connected to  the SWA and holds also for an infinite system. In 
Refs. \cite{PatrascioiuSeiler} arguments  were given that the model undergoes a 
freezing transition at non-zero temperature \cite{Patrascioiu02} with typical 
low-temperature configurations in a form of a gas of superinstantons \cite{Patrascioiu95}. 
Subsequently, an onset of the low-temperature 
QLRO phase is characterised by the power-law decay of the pair correlation function with 
an exponent $\eta=\eta^{XY}$ \cite{Patrascioiu01,PatrascioiuSeiler}. Note, that our result 
(\ref{2EtaXY}) does not contradict the estimate of \cite{Patrascioiu01}, since the 
former is valid at low temperatures, whereas the latter holds at the transition temperature.
Although our approach can not give a definite answer about the presence of QLRO phase in an 
infinite system, this question is still under discussion.

\begin{figure} [th]
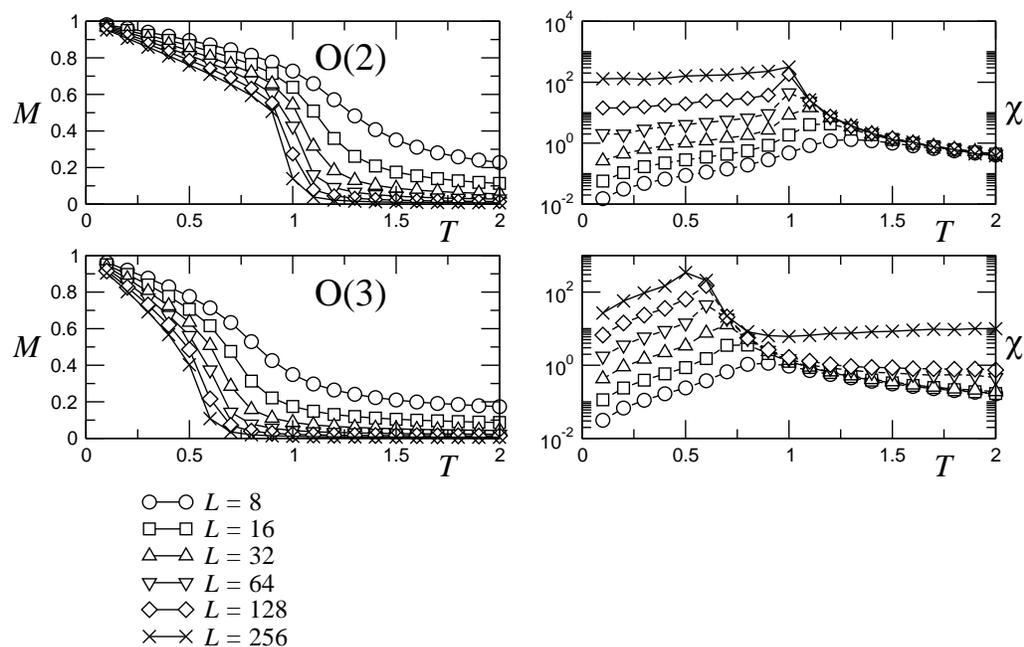

  \includegraphics[width=13.5cm]{Fig_Qtties_O2.eps}
  \includegraphics[width=13.5cm]{Fig_Qtties_O3.eps}
        \caption{Monte Carlo results for  $XY$ (to illustrate the well established case) 
	and Heisenberg models for finite two-dimensional lattices of different sizes.}
        \label{Fig2}  \vskip -0cm
\end{figure}

\begin{figure} [th]
  \centerline{\includegraphics[width=13.5cm]{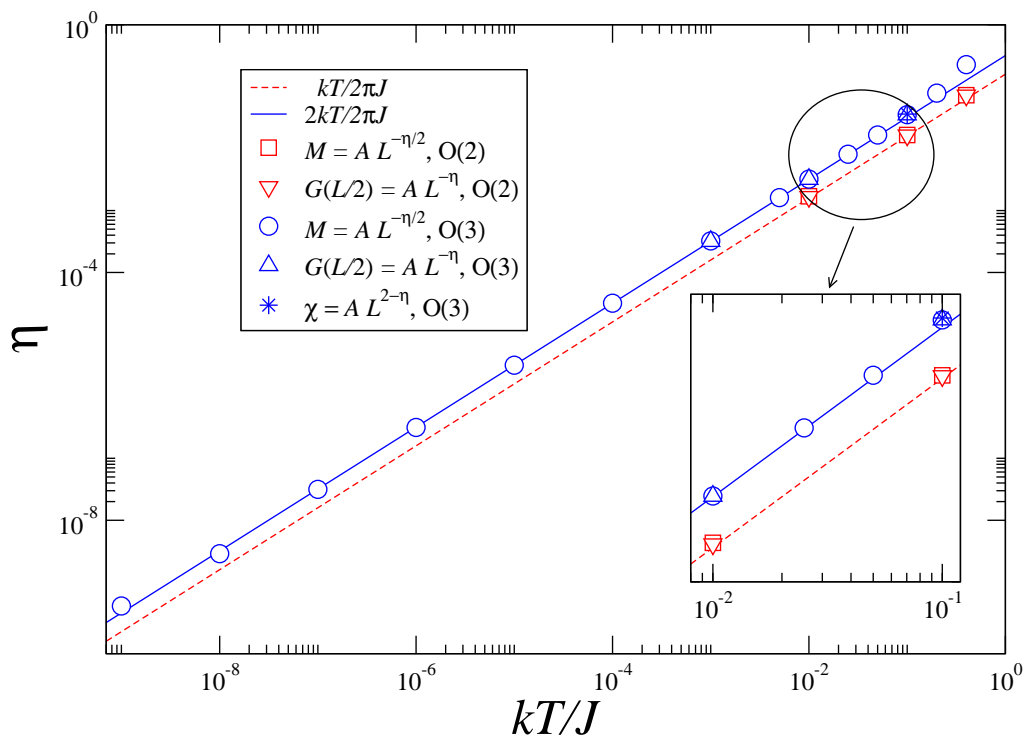}}
        \caption{Comparison between the exponents $\eta$ of the Heisenberg
                 model obtained from
                 the Monte Carlo simulations and from the analytic calculation
                 in the SWA. The dashed line presents $\eta^{XY}$. The inset 
	shows an increase of the scale to make the different symbols used 
	more visible.}
        \label{Fig3}  \vskip -0cm
\end{figure}

To verify our analytic results we have performed Monte Carlo
simulations of the Heisenberg spin model on lattices of different
sizes and at different temperatures. The Wolff's cluster algorithm
was used for this purpose \cite{Wolff89}. The exponent $\eta$ is obtained on the
base of three different observables, analysing the finite-size
scaling of the magnetisation, $M\sim L^{-\frac 12\eta(T)}$,
the pair correlation function, $G_2(L/2)\sim (L)^{-\eta(T)}$ and the
magnetic susceptibility, $\chi\sim L^{2-\eta(T)}$.
All three quantities are computed at different temperatures for varying
system sizes, giving access to a temperature-dependent exponent $\eta(T)$. 
Power-law scaling found for all three quantities $M$, $G_2$, and $\chi$
supports the presence of a QLRO phase found by analytic considerations.
Note that the lattice size in our
simulations changes from $N=8\times8$ to $N=256\times256$ for each
fixed temperature (Fig.~\ref{Fig2}) 
in order to obtain the finite-size scaling estimates of $\eta(T)$ shown in
Fig.~\ref{Fig3}. We cover the range of temperatures from $10^{-9}$ to the
order of 1. Thus in fact we observe both cases
$\frac{(R/a)^2}{4\beta J}\ll 1$ and $\frac{(R/a)^2}{4\beta J}\gg 1$.
However, comparing the analytic results and the outcome of our Monte
Carlo simulations in Fig.3, we see that $\eta=2\eta^{XY}$ fits the
Monte Carlo data over the whole range of temperatures except the
last several points (at the high temperature side of the
window shown in Fig.~\ref{Fig3}) 
which must indicate a transition to a non algebraic 
behaviour (see Refs.~\cite{Wolff2} and \cite{Patrascioiu01}).
Thus, $\eta$ defined by (\ref{1EtaXY}) have not
been observed in our computer simulations. The natural conclusion is
that when the condition $\frac{(R/a)^2}{4\beta J}\gg 1$ is reached
the temperature is not low enough to use the approach based on the
SWA and possibly is already close to the conjectured transition temperature
of the model (see Refs. \cite{Patrascioiu02}). Hence, the
question about a possibility to observe (\ref{1EtaXY}) in MC
simulations remains opened. But the important conclusion of the work
is that the result $\eta=2\eta^{XY}$ for the Heisenberg model in two
dimensions is reliable in a wide range of low temperatures up to
lattices $256\times256$.

\section*{Acknowledgements}

We acknowledge the CNRS-NAS exchange program and
Lo\"\i c Turban and Dragi Karevski for a discussion of the results.
We thank Erhard Seiler for useful correspondence.

\end{document}